\documentclass[fleqn,usenatbib]{mnras}

\usepackage{newtxtext,newtxmath}

\usepackage[T1]{fontenc}

\DeclareRobustCommand{\VAN}[3]{#2}
\let\VANthebibliography\thebibliography
\def\thebibliography{\DeclareRobustCommand{\VAN}[3]{##3}\VANthebibliography}

\usepackage{graphicx}	
\usepackage{amsmath}	

\title[Could kilomasers pinpoint supermassive stars?]{Could kilomasers pinpoint supermassive stars?}

\author[K. Nowak et al.]{
Katarzyna Nowak,$^{1}$\thanks{E-mail: k.nowak@herts.ac.uk}, Martin. G. H. Krause,$^{1}$ and Daniel Schaerer$^{2,}$$^{3}$
\\
$^{1}$Centre for Astrophysics Research, Department of Physics, Astronomy and Mathematics, University of Hertfordshire, College Lane, Hatfield AL10 9AB, UK \\
$^{2}$Observatoire de Genève, Université de Genève, Chemin Pegasi 51, 1290 Versoix, Switzerland \\
$^{3}$CNRS, IRAP, 14 Avenue E. Belin, 31400 Toulouse, France
}

\date{Accepted 2022 September 02. Received 2022 August 29; in original form 2022 July 19}

\pubyear{2022}

\begin{document}
\label{firstpage}
\pagerange{\pageref{firstpage}--\pageref{lastpage}}
\maketitle

\begin{abstract}

A strong nuclear kilomaser, W1, has been found in the nearby galaxy NGC 253, associated with a forming super star cluster. Kilomasers could arise from the accretion disc around supermassive stars (>10$^3$~M$_{\sun}$), hypothetical objects that have been proposed as polluters responsible for the chemical peculiarities in globular clusters. The supermassive stars would form via runaway collisions, simultaneously with the cluster. Their discs are perturbed by stellar flybys, inspiralling and colliding stars. This raises the question if an accretion disc would at all be able to survive in such a dynamic environment and mase water lines. We investigated what the predicted maser spectrum of such a disc would look like using 2D hydrodynamic simulations and compared this to the W1 kilomaser. We derived model maser spectra from the simulations by using a general maser model for appropriate disc temperatures. All our model discs survived. The model maser spectra for the most destructive case for the simulations of $M_{\star}$ = 1000~M$_{\sun}$ are a reasonable match with the W1 kilomaser spectrum in terms of scaling, flux values and some of the signal trends. Details in the spectrum suggest that a star of a few 1000~M$_{\sun}$ might fit even better, with 10,000~M$_{\sun}$ clearly giving too large velocities. Our investigations thus support the hypothesis that kilomasers could pinpoint supermassive stars.
\end{abstract}

\begin{keywords}
stars: abundances -- Galaxy: globular clusters: general -- galaxies: star clusters: general -- masers -- accretion, accretion discs -- hydrodynamics

\end{keywords}



\section{Introduction}

Kilomasers are much more luminous than masers from normal massive star formation sites in the Milky Way, but much less luminous than megamasers in Active Galactic Nuclei (AGN). Recently, a high-resolution spectrum became available of a nuclear kilomaser, W1, found in the nearby galaxy NGC 253 linked to a young massive cluster \citep[][previously observed by \citet{Brunthaler_2009} and \citet{Hofner_2006}]{Gorski_2018}. The association of this kilomaser and the ones found in NGC 4038 and NGC 4039 \citep{Brogan_2010} with young massive clusters is quite striking. \citet{Krause_2020} proposed that kilomasers could arise in accretion discs around supermassive stars (SMSs). \citet{Denissenkov_2014} and \citet{Gieles_2018} suggested that the SMSs, believed to be of a mass of at least 10$^3$~M$_{\sun}$, form via runaway collisions, simultaneously with a young massive cluster, usually in a very dense central region. They also proposed these so far hypothetical stars as polluters responsible for the chemical peculiarities in old globular clusters (GCs).

Globular clusters were traditionally believed to be simple, single population objects born in one coeval formation event with no internal chemical evolution. This view was completely revised in recent decades. For example, the 2$^\mathrm{{nd}}$ parameter problem \citep{Sandage_1967} arose in the 1960's when a sample of globular cluster colour-magnitude diagrams revealed that GCs with the same metallicity have different horizontal branch morphologies. GCs' with peculiar chemical compositions, including large variations in certain elements, were detected already in the 1970's \citep{Osborn_1971}, however their origins were then assigned to internal deep mixing processes as a consequence of the evolution of stars. 

Spectroscopic measurements in the 2000's showed that most Galactic and extra-galactic GCs demonstrate multiple sequences in the colour-magnitude diagram \citep{Anderson_2002, Bedin_2004, Piotto_2007, Piotto_2015, Milone_2012, Milone_2013, Martocchia_2018}, considered to be a result of a spread in helium abundance \citep[$\Delta Y$,][]{Norris_2004, D'Antona_2005,Charbonnel_2016, Chantereau_2016} and proving that globular clusters host multiple stellar populations \citep{Bastian_2018, Milone_2018}. Most globular clusters show no spread in iron abundance but display a similar maximum sodium enhancement. Helium abundance spreads vary ($\Delta Y$ $\lesssim 0.1$) from cluster to cluster, but are generally low \citep{Bastian_2015, Bastian_2018, Milone_2018, Lardo_2018}. GCs display large variations in light elements: Na-O, C-N and Mg-Al anticorrelations \citep{Denisenkov_1990, Langer_1993, Ventura_2001, Prantzos_2007, Gratton_2012, Charbonnel_2016, Prantzos_2017}. Still the most noticeable feature in most globular cluster is the Na-O anticorrelation. A hot-hydrogen burning environment is needed to vary those abundances with the concurrent p-capturing reactions of the CNO-cycle ($\gtrsim$ 20 MK), NeNa ($\gtrsim$ 45 MK) and MgAl ($\gtrsim$ 70 MK) chains leading to the rise of those anticorrelations \citep{Gratton_2012, Prantzos_2017}. 

Most models, seeking to explain the anomalies in globular clusters, refer to self-enrichment, where certain stars, polluters, within a cluster are capable of enriching other stars within the same cluster. It is also vital to include in the models how the observed amount of sodium can be produced and subsequently be accreted by the low mass proto-stars in the GCs; a requirement that multiple generation models struggle to meet. This issue is generally referred to as the 'mass budget problem' \citep{Bastian_2018, Gieles_2018}. In order to explain the above-mentioned anomalies three potential polluters have received a lot of attention in the literature: Asymptotic Giant Branch (AGB) stars \citep{Ventura_2001}, fast-rotating massive stars \citep[FRMS,][]{Decressin_2007b, Krause_2013} and supermassive stars \citep{Denissenkov_2014}. The nucleosynthesis of the first two proposed polluters does not correspond to the ones of GCs \citep{Bastian_2015, Prantzos_2017}. AGB stars display O-Na correlation instead of the anticorrelation observed \citep{Forestini_1997, Denissenkov_2003, Karakas_2007, Siess_2010, Ventura_2013, Doherty_2014, Renzini_2022}, furthermore it releases He-burning products, that are not widely detected in GCs \citep{Karakas_2006, Decressin_2009, Yong_2014}. FRMSs on the other hand could be able to produce Mg-Al anticorrelations\footnote{Assuming a significant increase of the $^{24}$Mg($p, \gamma$) reaction with respect to current predicted estimates \citep{Decressin_2007b}.} but simultaneously would show a strong He enrichment \citep{Decressin_2007b, Martins_2021}. The essential central temperature to activate the MgAl chain is reached by a supermassive star at the very beginning of its evolution when the abundance of He is still low \citep{Prantzos_2017}. Thereupon, in its early evolutionary phase, the H-burning products show agreement with various anticorrelations observed in the GCs \citep{Denissenkov_2014, Denissenkov_2015}. The SMS is assumed to be fully convective and therefore releases the material at the very beginning of the main sequence phase in a radiatively driven wind. The ejecta would then mix with star-forming gas that either accretes onto proto-stars or collapses to form stars independently \citep{Krause_2020}. The model of concurrent formation of proto-GCs and SMS proposed by \cite{Gieles_2018} provides the correct chemical patterns through the 'conveyor-belt' production of hot-H burning yields that also nicely solves the mass budget problem. The biggest disadvantage of the SMS model is arguably that no such objects are known to date \citep[cf.][]{Renzini_2022}.

Supermassive stars have also been proposed as a candidates for progenitors of supermassive black holes, but in this scenario they are expected to form through very rapid accumulation of gas \citep{Begelman_2010, Schleicher_2013, Haemmerle_2019}.

Supermassive stars would be extragalactic objects, surrounded by gas and dust as frequently observed for young, massive clusters \citep[e.g.,][]{Hollyhead_2015}, resulting in higher extinction. Effective temperature range between less than 10,000~K to $\approx$ 40,000~K \citep{Gieles_2019, Martins_2020}. For example for the hot case the star would be classified as blue, which would make such a star similar to normal massive stars, hard to resolve in far-away massive enough clusters and likely heavily absorbed in the gas-rich environment. This could be better for the case of a bloated and cooler star.  

Proto-stars in general have accretion disc and hence we would expect an SMS to also have one. An alternative method to detect them could therefore be MASER emission from any accretion disc around an SMS. GHz MASERs are often associated with massive star formation \citep{Ellingsen_2018, Billington_2019}. It is unclear, however, if an accretion disc would survive in an environment as dynamic as expected in the centre of a forming massive star cluster. A high rate of flybys, inspirals and collisions with the SMS might well inhibit the formation of the large molecular column with low velocity shear that are required for the emission of the maser line.

Flybys are known to be able to disrupt accretion discs down to the periastron radius of the flyby \citep{Clarke_1993, Cuello_2018, Vorobyov_2017,Vorobyov_2020}. For the SMS case, many stars will even collide with the SMS, but they should have much smaller mass than the SMS. Understanding the final outcome requires numerical simulation.

Here, we first summarise the observational arguments, why the W1 kilomaser is a candidate for an SMS accretion disc (Sect.~\ref{s:W1}), we then outline our hydrodynamics simulations (Sect.~\ref{s:model}) with which we are able to demonstrate that SMS are expected to maintain accretion discs and compare the results to observations of the W1 kilomaser (Sect.~\ref{s:results}). We discuss our findings in Sect.~\ref{s:discussion}, arguing that model details are consistent with an SMS mass of a few 1000~M$_{\sun}$ for the W1 system and summarise our conclusions in Sect.~\ref{s:conclusions}.

\section{The W1 kilomasers as a candidate for an accretion disc around a supermassive star}\label{s:W1}

The W1 kilomaser is an H$_2$O maser observed at 22.2~GHz. It is located in the starburst galaxy NGC~253 at a distance of 3.5~Mpc \citep{Rekola_2005} and associated with source 11 in \citet{Leroy_2018}. The still embedded super star cluster has an age of 1-2~Myr and a mass of 4$\times 10^5$~M$_{\sun}$ (compare their Table 2). Other kilomasers have likewise been associated with intense star formation, e.g., in the Antennae galaxies \citep[][and references therein]{Darling_2008}, and where spatial resolution was sufficient, they have been directly associated with super star clusters \citep{Brogan_2010}.

\begin{figure}
        \includegraphics[width=\columnwidth]{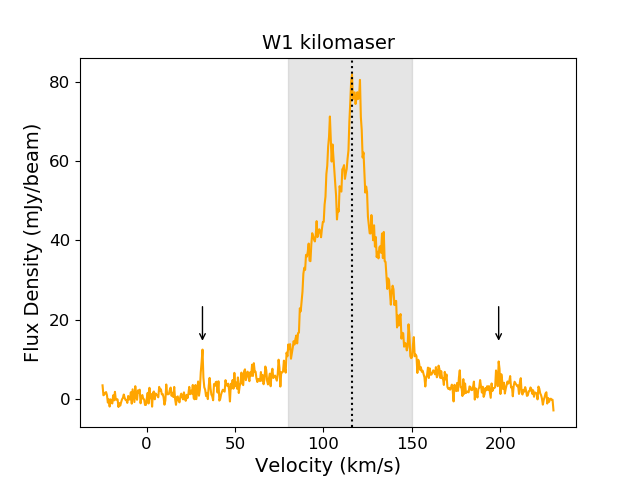}
        \caption{W1 kilomaser spectrum from~\protect\cite{Gorski_2018}. The dotted line indicates the systemic velocity of 116~km~s$^{-1}$. The low-velocity feature is highlighted by the shaded box. The black arrows on both sides of the systemic velocity point to the 'high-velocity' features at 31.6 and 199~km~s$^{-1}$.}
        \label{fig:W1_spectrum}
\end{figure}

The spectrum of the W1 kilomaser is shown in Fig.~\ref{fig:W1_spectrum}. It has three distinct line systems: the prominent one in systemic velocity of 116~km~s$^{-1}$ and two 'high-velocity' features on either side of the systemic velocity at substantially lower flux. If a maser shows two or the three corresponding lines (or line systems), it is called a clean disc maser \citep{Pesce_2015}, as typically observed for AGN megamasers. The spectrum of W1 looks like a disc maser spectrum, but is much weaker than the typical AGN megamasers, about two orders of magnitude lower in luminosity.

Some extragalactic kilomasers in super star clusters have been compared to the Galactic high mass star forming region W49N, located roughly 11.1~kpc away \citep{Gwinn_1992}. This region produces a large number of highly variable 22~GHz H$_2$O maser spots with the total luminosity of $\sim$ 1~L$_{\sun}$ \citep{Zhang_2013, Shakhvorostova_2019, Volvach_2020}. The spectrum consists of 316 individual narrow lines between velocities -352.1 and 375.5~km~s$^{-1}$ \citep{McGrath_2004}, but in all cases, the extragalactic kilomasers appear to have a more peaked and narrower spectrum. The best available kilomaser spectrum is probably from W1 \citep{Gorski_2018}. The spectrum of W1 is clearly different from the one of W49N and instead consistent with the clean disc maser spectrum as argued above.

\begin{figure}
        \includegraphics[width=\columnwidth]{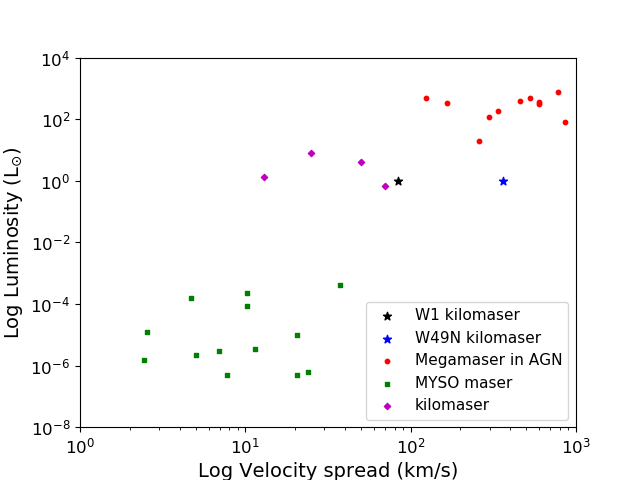}
        \caption{Luminosity versus velocity spread plot for water masers from different sources. Red circles indicate megamasers from AGN sources, green one denotes stellar masers from massive young stellar objects (MYSOs), purple circles show kilomasers, the blue star is the W49N and the black star is the W1 kilomaser. See Table~\ref{tab:maserclassification} for references.}
        \label{fig:Maser_spread}
\end{figure}

Figure~\ref{fig:Maser_spread} compares velocity spread versus luminosity of the W1 kilomaser with water masers from different types of sources, including AGNs, massive young stellar objects (MYSOs) and other kilomasers, associated with other forming super star clusters and W49N. The spread in velocity space is a few hundred km~s$^{-1}$ for megamasers and W49N, whereas for the W1 kilomaser, it is around 80 km~s$^{-1}$. We have extracted this data from various references displayed in Table~\ref{tab:maserclassification}. The relationship between luminosity and velocity spread is super-linear, and extragalactic kilomasers including the W1 kilomaser are in the middle between massive YSOs and AGNs. Hence, it would seem reasonable that kilomasers belong to an object class that is in between normal massive stars $\lesssim 10^2$~M$_{\sun}$ and AGNs $\gtrsim 10^6$~M$_{\sun}$.

\begin{table*}
\centering
	\caption{H$_2$O masers in different astronomical sources. Kilomasers have their host galaxy indicated in the brackets in the first column. All the stellar masers are from massive young stellar objects (MYSOs), whilst megamasers are from AGN sources.}
	\label{tab:maserclassification}
		\begin{tabular}{lcccr}
			\hline
			Name of the source & Luminosity & Max vel. spread & Classification & References \\
			 & (L$_{\sun}$) & (km~s$^{-1}$) & & \\
			\hline
			NGC 4258 & 80 & 890 & Megamaser & \citep{Greenhill_1996} \\
			NGC 3079 & 500 & 126 & Megamaser & \citep{Yamauchi_2004}\\
			Circinus & 20 & 260 & Megamaser & \citep{Greenhill_2003}\\
			UGC 3789 & 370 & 725 & Megamaser & \citep{Reid_2009}\\
			NGC 2960 (Mrk 1419) & 400 & 465 & Megamaser & \citep{Henkel_2002, Kuo_2011}\\
			IC 2560 & 122 & 323 & Megamaser & \citep{Ishihara_2001}\\
			NGC 3393 & 320 & 601 & Megamaser & \citep{Kondratko_2008}\\
			J0437+2456 & 178 & 338.85 & Megamaser & \citep{Gao_2017}\\
			NGC 6323 & 500 & 600 & Megamaser & \citep{Kuo_2011, Kuo_2014}\\
			NGC 6926 & 340 & 200 & Megamaser & \citep{Sato_2005}\\
			UGC 6093 & 770 & 801.28 & Megamaser & \citep{Zhao_2018}\\
			W1 (NGC 253) & 1.02 & 84.4 & Kilomaser & \citep{Gorski_2018} \\
			W49N  & 1 & 363.8 & Kilomaser & \citep{McGrath_2004} \\
			H$_2$O-East (NGC 4038/NGC 4039) & 1.3 & 13 & Kilomaser & \citep{Brogan_2010}\\
			H$_2$O-SE (NGC 4038/NGC 4039) & 4.1 & 50 & Kilomaser & \citep{Brogan_2010} \\
			H$_2$O-West (NGC 4038/NGC 4039) & 7.7 & 25 & Kilomaser & \citep{Brogan_2010}\\
			He 2-10 & 0.68 & 70 & Kilomaser & \citep{Darling_2008}\\
			G229.5711+00.1525 & 8.5$\times$10$^{-5}$ & 10.25 & Stellar maser &\citep{Urquhart_2011}\\
			G220.4587-00.6081 & 2.29$\times$10$^{-6}$ &5 & Stellar maser & \citep{Urquhart_2011}\\
			G083.7071+03.2817 & 4.79$\times$10$^{-7}$ & 7.75& Stellar maser &\citep{Urquhart_2011}\\
			G080.8624+00.3827 & 2.88$\times$10$^{-6}$ & 6.9 & Stellar maser & \citep{Urquhart_2011}\\
			G081.8652+00.7800 & 4.27$\times$10$^{-4}$ & 37.5& Stellar maser & \citep{Urquhart_2011}\\
			G081.7131+00.5792 & 3.55$\times$10$^{-6}$ & 11.5& Stellar maser &\citep{Urquhart_2011}\\
			G081.7624+00.5916 & 6.03$\times$10$^{-7}$ & 23.85& Stellar maser &\citep{Urquhart_2011}\\
			G084.1940+01.4388 & 5.13$\times$10$^{-7}$ & 20.6& Stellar maser &\citep{Urquhart_2011}\\
			G084.9505-00.6910 & 1.55$\times$10$^{-6}$ & 2.45& Stellar maser &\citep{Urquhart_2011}\\
			G095.0531+03.9724 & 1.62$\times$10$^{-4}$ & 4.7& Stellar maser &\citep{Urquhart_2011}\\
			G094.2615-00.4116 & 9.55$\times$10$^{-6}$ & 20.6& Stellar maser &\citep{Urquhart_2011}\\
			G094.4637-00.8043 & 1.26$\times$10$^{-5}$ & 2.55& Stellar maser &\citep{Urquhart_2011}\\
			G094.6028-01.7966 & 2.29$\times$10$^{-4}$ & 10.2& Stellar maser &\citep{Urquhart_2011}\\
			\hline
		\end{tabular}
\end{table*}

\section{An accretion disc model for collisionally maintained supermassive stars}\label{s:model}
In the following, we first derive constraints on the model parameters from relevant observations and then describe the setup of the hydrodynamic simulations used to test the stability of an SMS accretion disc in the dynamical environment in the centre of a massive young cluster. Our aim is to model the collisionally pumped 'high-velocity' features of the W1 kilomaser, which contain information on the mass of the central object. We do not attempt to model radiatively pumped features near systemic velocity.

\subsection{General model}
The similarity of the W1 maser spectrum to AGN megamaser discs suggests that it may be useful to apply the physics learnt from AGN to kilomasers. The 'high-velocity' features can be used to calculate the radius of the disc, as they represent emission closer to the outer edges of the disc in AGNs. Hence assuming this also applies to the disc around a supermassive star and the kilomaser it produces, we take the velocity spread out to both 'high-velocity' features in the W1 kilomaser to be $\Delta\nu$ = 83~km~s$^{-1}$. This measurement was performed at a digital copy of the original data kindly provided by Mark Gorski. We assume that all the high-velocity maser spots are on circular Keplerian orbits on the midline of the disc \citep{Pesce_2015}. Keplerian dynamics then gives the radius of the maser spot as:
\begin{equation}
    R_\mathrm{out} = \frac{GM_{\mathrm{{\star}}}}{(\Delta \nu)^2}.
    \label{eq:outerrad}
\end{equation}
where $G$ is gravitational constant and M$_\mathrm{\star}$ is mass of the central star. For a 1000~M$_{\sun}$ star, we would have $R_\mathrm{out} \sim$ 129~au ($\sim$ 6.2$\times$10$^{-4}$~pc). The high-velocity maser spots in the W1 kilomaser might be located in the closest area to the inner disc that can mase the water line and not necessarily in the vicinity of the outer edges of the disc. We compare this to candidate discs of high-mass protostars by \citet{Cesaroni_2006}. One of the candidates listed, IRAS~20126+4104, is the best-studied case of a Keplerian accretion disc around high mass star. The disc's mass is about 4~M$_{\sun}$ with a radius of 1600~au around a massive (7~M$_{\sun}$) YSO \citep{Cesaroni_2005}. The most common value of radius of the disc is 500~au and some values are as high as 20,000~au\footnote{Although the authors argue that those high values of radius would suggest it to be a massive rotating structure called 'toroid', rather than an accretion disc around a single star. 'Toroids' are believed to host a stellar cluster and might not be in equilibrium, whilst accretion discs are in Keplerian orbits around their central stars \citep{Cesaroni_2006}.} \citep[see also][]{Ilee_2018, Sanna_2021, Williams_2022}.

Taking all the above arguments into consideration we simulated accretion discs with the inner and outer radius of $R_\mathrm{in}$ = 10~au and $R_\mathrm{out}$ = 500~au or 1000~au. We stress that the inner radius chosen is not meant to imply that the disc would not continue much further towards the star. The disc in this region would, however, likely to be too hot to allow the formation of maser spots (compare below). 

Masses of accretion discs vary substantially between objects and do not scale with the radius of the disc or the mass of the central star \citep{Cesaroni_2006}. We decided for 1 per cent (similar to the flyby simulations of \citet{Cuello_2018}) and 10 per cent of $M_{\star}$.

Following \citet{Cuello_2018}, we assumed a power law surface density at the start of the calculation:
\begin{equation}
    \Sigma = \Sigma_0\left(\frac{R}{R_\mathrm{in}}\right)^{-1}
    \label{eq:sigma}
\end{equation}
\begin{equation}
    \Sigma_0 = \frac{M_\mathrm{disc}}{2\pi R_\mathrm{in}(R_\mathrm{out}-R_\mathrm{in})}
\end{equation}
giving a height varying with radius as:
\begin{equation}
    H = 0.05 R_\mathrm{in}\left(\frac{R}{R_\mathrm{in}}\right)^{1.25}.
\end{equation}
The density of the disc is then given by:
\begin{equation}
    \rho = \frac{\Sigma}{2 H} = \frac{\Sigma_0 R_\mathrm{in}}{2R H},
\end{equation}
where we have used Eq.~\ref{eq:sigma} for the second equality. The density outside the disc is set to 10$^{-20}$~g~cm$^{-3}$.

Temperature in discs around massive stars is known from observations and radiative transfer calculations to be similar to the stellar surface temperature close to the star and then drops off with a power-law index between -0.5 and -1, where observations seem to indicate an index closer to -0.5 \citep{Lesniak_2011,Akiyama_2013,Vural_2014, Qiao_2022}. Supermassive stars were modelled to have temperatures between about 10~kK to 40~kK and radii between 10$^2$ and few times 10$^3$ \citep{Gieles_2018, Gieles_2019}. Our resulting estimate for the mass of the hypothetical SMS in the W1 system depends linearly on the radius of the maser spots and hence linearly on the temperature and the radius of the SMS. We therefore chose a conservative estimate of the effective temperature, $T_\mathrm{eff}$, of the star of 9~kK and a radius of 100~R$_{\sun}$. We hence set the temperature as a function of radius to a scaled up law from \citet{Cuello_2018}:
\begin{equation}
    T = 1920\, \mathrm{K}\left(\frac{R}{R_\mathrm{in}}\right)^{-1/2},
    \label{eq:temperature}
\end{equation}
where the first value is the temperature at the inner edge of the disc in our simulations. Up to around 100~au radius the temperature is higher than 300~K and hence allows for the water lines to be mased.
A locally isothermal equation of state is used with the temperature as defined by Eq.~\ref{eq:temperature}. 

The angular velocity is calculated from the below equation of radial hydrostatic equilibrium equation, which is preserved by the balance of gravitational acceleration with centrifugal acceleration and the pressure gradient:
\begin{equation}
    \frac{1}{\rho}\frac{\partial p}{\partial R} = \Omega^2 R - \frac{GM_\mathrm{\star}}{R^2},
\end{equation}
where pressure is $p = \frac{\rho k_\mathrm{B} T}{\mu m_\mathrm{p}}$, with Boltzmann constant $k_\mathrm{B}$, mean molecular weight $\mu$ = 2.35 \citep{Kimura_2016}, $m_\mathrm{p}$ as a mass of a proton and $T$, temperature expression mentioned above. Therefore the angular velocity for the SMS disc is:
\begin{equation}
    \Omega^2 = \frac{GM_{\star}}{R^3} - 0.5 \frac{1920 \mathrm{K}k_\mathrm{B}}{ \mu m_\mathrm{p}R}\left(\frac{R}{R_\mathrm{in}}\right)^{-0.5}.
    \label{eq:angularvel}
\end{equation}

According to \citet{Clarke_1993} the most destructive encounter, where the disc looses around 50 per cent of its mass (if the ratio of the outer edge of the disc and the periastron distance, $R_\mathrm{peri}$, is 0.8) is prograde and coplanar, whilst the retrograde and coplanar one has almost no impact on the disc. In the scenario, where orbital and disc planes are close to orthogonal, the perturber plunges through the disc without having any previous interaction with. Therefore the disc particles located within periastron remain bound and sustain their alignment, while the rest of the material is transferred out on to orbits inclined to the original orbital plane. Therefore we will only consider the case where the encounter affects the disc in the most destructive manner, namely prograde and coplanar (inclination of the orbit of 0${\degr}$ angle).

Flyby simulations are typically done with similar masses of the perturbers and the star with the accretion disc. For the case we consider expected perturber mass is much lower than the expected mass of the SMS. However, there are many perturbers expected which can add up to a mass comparable to the mass of the SMS. The mass of the perturbers is estimated from Figure 3 in \citet{Gieles_2018}. We took the 'metal-poor' case with the number of stars $N$ = 10$^7$. At the time when $m_\mathrm{sms}$ is equal 10$^3$~M$_{\sun}$, the total mass of all the stars in the cluster is 10$^{6.2}$~M$_{\sun}$. Thus the average mass of the pertuber is $M_\mathrm{pert} \sim $ 0.2~M$_{\sun}$. 

Using the same plot from \citet{Gieles_2018} we estimated the number of collisions and their frequency. We took the initial mass in stars $M_{0}$ = 10$^6$~M$_{\sun}$, the cluster half-mass radius of $R_\mathrm{h0}$ $\simeq$ 2.3~pc and the initial mass of each star, $m_0$ $\simeq$ 0.1~M$_{\sun}$. The accretion factor, $a$, for the mass in the cluster i.e. current mass of the stars (when $M_{\mathrm{SMS}}$ = 1000~M$_{\sun}$) divided by the initial mass of the stars, is $\sim$ 2. The value higher than that would corresponds to two-body relaxation becoming more important \citep{Gieles_2018}. Therefore the total mass of the stars in the cluster is:
\begin{equation}
    M = a M_\mathrm{0},
\end{equation}
which gives the value of 2$\times$10$^6$~M$_{\sun}$. The radius, in which the mass of the stars is contained is $r_{0}$ = a$^{-3}$   $R_\mathrm{h0}$ = 0.085~pc (17500~au). We are only interested in the region of $r_\mathrm{i}$ = 1000~au, as this is the highest value of the accretion disc's radius we are simulating. The density of stars:
\begin{equation}
    n = \frac{3M}{4\pi r_\mathrm{0}^3 a m_\mathrm{0}},
\end{equation}
gives 4.45 $\times$ 10$^{-7}$ stars per (au)$^3$. Hence the number of stars, $N$, in the region of interest, $r_{i}$:
\begin{equation}
    N = \frac{4\pi r_\mathrm{i}^3 n}{3},
\end{equation}
is 1864. The crossing times is defined as $t_\mathrm{cross}$ = 2$r_\mathrm{i}$/$V_\mathrm{rms}$, where $V_\mathrm{rms}$, is the one-dimensional velocity dispersion of the cluster:
\begin{equation}
    V_\mathrm{rms} \simeq \sqrt{\frac{GM}{6R_\mathrm{ho}}} = 159~\mathrm{km~s}^{-1}.
\end{equation}
Therefore, the crossing time, $t_\mathrm{cross}$, is 60~years. Applying the values calculated above, the flyby rate = $N/t_\mathrm{cross}$  $\simeq$ 30 per year. We are only interested in the collisions that occur in the plane of the disc, 10 per cent of the solid angle, and only in the prograde ones. Therefore the flyby rate is around 1 per year to the order of the magnitude. 

Mark Gieles (private communication) has kindly provided his estimate of the collision rate of 0.025 per year, and for the ones that occur in the plane of the disc is equal to 0.0025 flybys per year, therefore around 1 per 400 years. This is certainly a lower limit, as it only takes into account the actual hits, not the near misses that can still impact the accretion disc significantly. Guided by these constraints, we simulated flyby rates from 10$^{-2}$ yr$^{-1}$ to 1 yr$^{-1}$

The initial position of each perturber is 10 times the value of the SMS disc's outer radius, $R_\mathrm{out}$. As mentioned in \citet{Cuello_2018}, by setting this initial value it removes any artificial effects caused by a sudden introduction of the perturber close to the disc. The perturber is entered in the negative $y$ and $x$ direction, and leaves the grid towards negative $y$-direction but positive $x$-direction, allowing for the prograde encounter. The initial $y$ and $x$ values of the perturber are calculated as follows:
\begin{equation}
    y_\mathrm{i} = -10 R_\mathrm{out}
\end{equation}
\begin{equation}
    x_\mathrm{i} = -2 R_\mathrm{peri}\sqrt{1-\frac{y_\mathrm{i}}{R_\mathrm{peri}}}
\end{equation}
In order for the perturber to follow a parabolic orbit we use Barker's equation:
\begin{equation} \label{eq:barker}
    \Delta t = \sqrt{\frac{2R_\mathrm{peri}^3}{GM_\mathrm{t}}}\left( D_\mathrm{f} + \frac{1}{3}D_\mathrm{f}^3-D_\mathrm{i} -\frac{1}{3}D_\mathrm{i}^3\right)
\end{equation}
where $M_\mathrm{t}$ is the sum of the mass of the central star, $M_{\star}$ and the mass of the perturber, $M_\mathrm{pert}$. 
Subscripts 'i' and 'f' indicate the initial and final positions of the perturber,
\begin{equation} \label{eq:D}
    D_\mathrm{i,f} = \tan\left(\frac{\nu_\mathrm{i,f}}{2}\right)
\end{equation}
and true anomaly is:
\begin{equation} \label{eq:nu}
    \nu_\mathrm{i,f} = \arctan\left(\frac{x_\mathrm{i,f}}{y_\mathrm{i,f}}\right).
\end{equation}
We use equations~\ref{eq:barker}, ~\ref{eq:D} and ~\ref{eq:nu} to calculate the position of the perturber as a function of time. According to \citet{Meire_1985} the most practical solution to the Barker's equation follows the below calculations:
\begin{equation}
    A = \frac{3}{2}\sqrt{\frac{2GM_{\star}}{2R_\mathrm{peri}^3}}\left( t-t_\mathrm{0}\right)
\end{equation}
\begin{equation}
    B = \left( A + \sqrt{A^2+1} \right) ^{\frac{1}{3}}.
\end{equation}
Hence the true anomaly of the perturber's orbit is:
\begin{equation}
    \nu = 2\arctan\left( B-\frac{1}{B}\right)
\end{equation}
and its radius is determined as follows:
\begin{equation}
    R = \frac{2 R_\mathrm{peri}}{1+\cos{\nu}}
\end{equation}
for each perturber. Flyby rate as well as the time of the simulation allows for the number of perturbers to be calculated i.e. for 5000 years of simulation time and a flyby rate of 1 per year, we included 5000 perturbers. Periastron distance values are generated using random numbers up to the maximum value of periastron distance set as $R_\mathrm{peri}$ = 1.1~$R_\mathrm{out}$. The start time of the perturbers is also random. That ensures the perturbers are spread out in time according to the chosen flyby rate. To ensure perturbers are coming from different angles, without changing the starting values of $y_\mathrm{i}$ and $x_\mathrm{i}$, we rotated the initial perturber positions using the rotation matrix:
\begin{equation}
R = 
 \begin{bmatrix}
x_\mathrm{p} \cos{\theta}  - y_\mathrm{p} \sin{\theta} \\
x_\mathrm{p} \sin{\theta}  + y_\mathrm{p} \cos{\theta}
\end{bmatrix},   
\end{equation}
where $\theta$ is the orbital angle and its values are generated using random numbers between 0 and 2$\pi$. 

We simulated only prograde encounters, hence only half of the encounters calculated above, as they are more destructive than the retrograde ones.  

We defined direct stellar collision of the perturber with the central star when the periastron distances, $R_\mathrm{peri}$ $\leq$ 10~au. In this instance the position of perturber, $x_\mathrm{p}$ and $y_\mathrm{p}$, as well as its radius, $R$, are set to 0. The mass of the perturber, $M_\mathrm{pert}$ is then added to the mass of the central star, $M_{\star}$. Direct collisions are purely determined by the values of $R_\mathrm{peri}$ for each individual stellar perturber. Therefore their number is random. For the case of flyby rate of one stellar perturber per 100 years there is no direct collision ($R_\mathrm{peri}$ $\leq$ 10~au).

Similarly to the work of \citet{Vorobyov_2017, Vorobyov_2020} the perturber potentials are smoothed \citep{Klahr_2006}, such that the total potential is given by:
\begin{equation}\label{eq:gravit}
    \Phi = \Phi_{\star} + \Phi_\mathrm{p} = -\frac{GM_{\star}}{r} + \Phi_\mathrm{p},
\end{equation}
where the cubic perturber's potential $\Phi_\mathrm{p}$ is defined as:
\begin{equation}\label{eq:rsm}
   \Phi_\mathrm{p} =
    \begin{cases}
      -\frac{m_\mathrm{p}G}{d}\left[\left(\frac{d}{r_\mathrm{sm}}\right)^4 - 2\left(\frac{d}{r_\mathrm{sm}}\right)^3 + 2\frac{d}{r_\mathrm{sm}}\right] & \text{for $d$} \leqslant r_\mathrm{sm}\\
      -\frac{m_\mathrm{p}G}{d} & \text{for $d$} > r_\mathrm{sm},
    \end{cases}
\end{equation}
with the distance from the perturber $d$ and the smoothing radius $r_\mathrm{sm}$ = 0.13$r$, where $r$ is the distance of the given cell from the SMS. The top term in the above equation has an additional term in the square brackets, which artificially reduces the strength of the potential inside $r_\mathrm{sm}$. Since the resolution is limited and the region is unresolved, the value of the acceleration would be wrong. The application of the above term allows for the correct calculation of the accelerations on the scales that can be resolved, minimising the errors.

\subsection{Computational setup}
To simulate the accretion disc of the supermassive star, the finite volume fluid dynamics code {\sc PLUTO} is used \citep{Mignone_2007}. It has been designed to integrate a system of conservation laws given by:
\begin{equation}
    \frac{\partial\bmath{U}}{\partial t} = -\bigtriangledown \cdot \bmath{T}\left(\bmath{U}\right) + \bmath{S}\left(\bmath{U}\right)
    \label{conserlaw}
\end{equation}
where the state vector $\bmath{U}$ represents a set of conservative variables, $\bmath{T}\left(\bmath{U}\right)$ describes fluxes of each component of the state vector and $\bmath{S}\left(\bmath{U}\right)$ represents the source terms.

For the case of ordinary hydrodynamics, Eq.~\ref{conserlaw} reduces to the following Euler equations:
\begin{equation}
    \frac{\partial\rho}{\partial t} + \bigtriangledown \cdot \left(\rho \bmath{v}\right) = 0 
\end{equation}
\begin{equation}
    \frac{\partial\bmath{\rho\bmath{v}}}{\partial t} + \bigtriangledown \cdot \left( \rho\bmath{v}\bmath{v} + p\bmath{I}\right) = -\rho\bigtriangledown\Phi
\end{equation}
where $\rho$ is the mass density, $\bmath{v}$ is the velocity, $p$ is the thermal pressure and the gravitational potential is $\Phi$.

The equation of state provides closure:
\begin{equation}
    p = \rho c_\mathrm{s}^2,
\end{equation}
where the sound speed is a function of radius $c_\mathrm{s}(r)$, as given by Eq.~\ref{eq:temperature}.

We run {\sc PLUTO} on the University of Hertfordshire high performance computing cluster (UHHPC). The code is set up to work with non-dimensional units and hence the dimensionalisation to c.g.s. units is essential for the simulation. We tested the simulation setup for a single parabolic stellar flyby of both perturber and central star of mass 1~M$_{\sun}$ and successfully reproduced the results published by \citet{Cuello_2018}. The HD equations are solved in two dimensions in spherical coordinates. The spatial integration is performed using linear reconstruction of the fluxes to the cell boundaries and time evolution is computed using a second order Runge-Kutta scheme. The Harten-Lax-van Leer-Contact discontinuity (HLLC) Riemann solver is used to solve the numerical fluxes. The radial grid is defined from 10~au to 5000~au for $R_\mathrm{out}$ = 500~au and from 10~au to 10,000~au for $R_\mathrm{out}$ = 1000~au using a logarithmic scale. The azimuthal, uniform grid is set from 0 to 2$\pi$ with periodic boundary. We used a resolution of 50x100 grid elements for radial and azimuthal direction respectively. We performed a set of simulations, varying different parameters outlined in Table~\ref{tab:SMS_simulationset}.

\begin{table}
\centering
	\caption{Set of simulations of accretion disc around the supermassive star, varying its mass, mass and outer radius of the accretion disc, periastron distance and flyby rate.}
	\label{tab:SMS_simulationset}
		\begin{tabular}{lcccr}
			\hline
			Label&$M_{\mathrm{SMS}}$ & $R_\mathrm{out}$ & $M_\mathrm{disc}$ & flyby rate  \\
			&(M$_{\sun}$) &(au)& ($M_{\mathrm{SMS}}$) & (year$^{-1}$) \\
			\hline
			\hline
			SMS-mrd-hf &1000 & 500 & 0.01 & 1  \\
			SMS-mrd-mf &1000 & 500 & 0.01 & 10$^{-1}$\\
			SMS-mrd-lf &1000 & 500 & 0.01 & 10$^{-2}$\\
			\hline
			SMS-mrD-hf &1000 & 500 & 0.1 & 1 \\
			SMS-mrD-mf &1000 & 500 & 0.1 & 10$^{-1}$\\
			SMS-mrD-lf &1000 & 500 & 0.1 & 10$^{-2}$\\
			\hline
			SMS-mRd-hf &1000 & 1000 & 0.01 & 1 \\
			SMS-mRd-mf &1000 & 1000 & 0.01 & 10$^{-1}$\\
			SMS-mRd-lf &1000 & 1000 & 0.01 & 10$^{-2}$\\
			\hline
			SMS-mRD-hf &1000 & 1000 & 0.1 & 1  \\
			SMS-mRD-mf &1000 & 1000 & 0.1 & 10$^{-1}$\\
			SMS-mRD-lf &1000 & 1000 & 0.1 & 10$^{-2}$\\
			\hline
			SMS-Mrd-hf &10000 & 500 & 0.01 & 1  \\
			SMS-Mrd-mf &10000 & 500 & 0.01 & 10$^{-1}$\\
			SMS-Mrd-lf &10000 & 500 & 0.01 & 10$^{-2}$\\
			\hline
			SMS-MrD-hf &10000 & 500 & 0.1 & 1  \\
			SMS-MrD-mf &10000 & 500 & 0.1 & 10$^{-1}$\\
			SMS-MrD-lf &10000 & 500 & 0.1 & 10$^{-2}$\\
			\hline
			SMS-MRd-hf &10000 & 1000 & 0.01 & 1  \\
			SMS-MRd-mf &10000 & 1000 & 0.01 & 10$^{-1}$\\
			SMS-MRd-lf &10000 & 1000 & 0.01 & 10$^{-2}$\\
			\hline
			SMS-MRD-hf &10000 & 1000 & 0.1 & 1  \\
			SMS-MRD-mf &10000 & 1000 & 0.1 & 10$^{-1}$\\
			SMS-MRD-lf &10000 & 1000 & 0.1 & 10$^{-2}$\\
			\hline
			\hline
		\end{tabular}
\end{table}

For the simulation runs SMS-mrd and SMS-Mrd the orbital period at our typical outer disc radius of 500~au is 355 and 112 years, respectively. We run the simulations without any perturber for 5000 years, which allows the disc to settle from initial high-amplitude oscillations. Smaller oscillations are ignored since the high number of stellar flybys and collisions will have a more substantial impact on the disc than the oscillations itself. After this initial period the actual run-time of the simulations starts, where the shortest run is around 1500 years for simulations with $M_{\mathrm{SMS}}$ = 10,000~M$_{\star}$, allowing the accretion disc to fully rotate at least 13 times.

\subsection{Model maser spectrum}
We derived the model maser spectra and plotted it together with the W1 kilomaser from \citet{Gorski_2018}, kindly provided in electronic form by Mark Gorski, for direct comparison. The systemic velocity of W1 in NGC~253 is 116~km~s$^{-1}$. For the integration along the line of sight, we map the simulation output on to a Cartesian grid and assume the source to be observed edge-on. Each of these Cartesian cells has one particular velocity and is regarded as a velocity coherent maser region or a part thereof with length d$y$. We need to adapt values for the linewidth and the maser spot size to determine the spectrum. The values for these can vary: for extragalactic kilomasers those values have not been measured, for AGN, a typical value is 3~km~s$^{-1}$ \citep{Kartje_1999} and for the Galactic kilomaser W49N is 0.5-1.9~km~s$^{-1}$ \citep{Liljestrom_2000}. As an example we take 1~km~s$^{-1}$, as with this value we already obtain enough flux to reproduce the observations, with a higher linewidth we get even higher fluxes. For regions along the line of sight that have the same velocity within a fiducial velocity coherence range of 1~km~s$^{-1}$, d$y$ is multiplied by the number of these regions, $n$, to obtain the total length of the velocity coherent region. The resolution of the grid is approximated based on the maser spot size of $\sim$ 10~au which is also roughly consistent with the marginally resolved maser spots in W49N from \citet{Zhang_2013}. If we decrease the spot size, the flux increases. 

Fluxes are calculated using the flux equation to model AGN maser discs derived by \citet{Kartje_1999}:
\begin{equation}
    F = 4.7\times10^{17}a_{10}\left(\frac{l}{D}\right)^2 \mathrm{Jy},
    \label{Kartje_fluxes}
\end{equation}
where $a_{10}$ = $a$/10, and $a$ defines the effective aspect ratio:
\begin{equation}
    a = \frac{l}{\sqrt{A/\pi}}.
    \label{aspectratio}
\end{equation}
For \emph{a} larger than a few the maser is saturated. We assumed $a_{10}$ = 1. The half-length of the maser emission region that lies in a disc along the line of sight is defined as $l$, $D$ is the distance to the source with the observed area $A$. That results in Eq.~\ref{Kartje_fluxes}, reducing to:
\begin{equation}
    F = 4.7\times10^{17}\left(\frac{n \mathrm{d}y}{2D}\right)^2 \mathrm{Jy},
\end{equation}
where $D$ = 3.5~Mpc is the distance to the starburst galaxy NGC~253. 

The 22~GHz H$_2$O line requires dense gas of at least 10$^7$~cm$^{-3}$ and temperatures larger than 300~K \citep{Gorski_2018}, hence we restrict the density for masing regions accordingly. The temperature defined by Eq.~\ref{eq:temperature} is also restricted between the already mentioned minimum of 300~K up to 1500~K, as to allow the gas to remain molecular \citep{Kartje_1999}. Those parameters are required for the water to be mased if it is collisionally pumped.

Since we only model collisionally pumped masers, the model spectra do not include the 'low-velocity' features, which are radiatively pumped by the the background infrared radiation, usually by the central object. For the purpose of this study, we only focus on the 'high-velocity' features in maser spectra. The model maser spectrum shows some residual positive flux at systemic velocity for each simulation run. This is the result of flux being calculated for each area in the disc that meet the density and temperature requirements for maser emission in the line of sight, thus computing the emission as a ring in the disc. According to \citet{Elitzur_1991} around 51 per cent of the maser emission is lost through the sides, which would actually result in such a positive flux around zero velocity. As the resulting fluxes are low, we make no attempt to model the sideways emission accurately. This approximation needs to be taken into consideration when interpreting the results.

\section{Results}\label{s:results}

\subsection{Disc stability for a 1000~M$_{\sun}$ star} 
\begin{figure*}
    \includegraphics[width=\columnwidth]{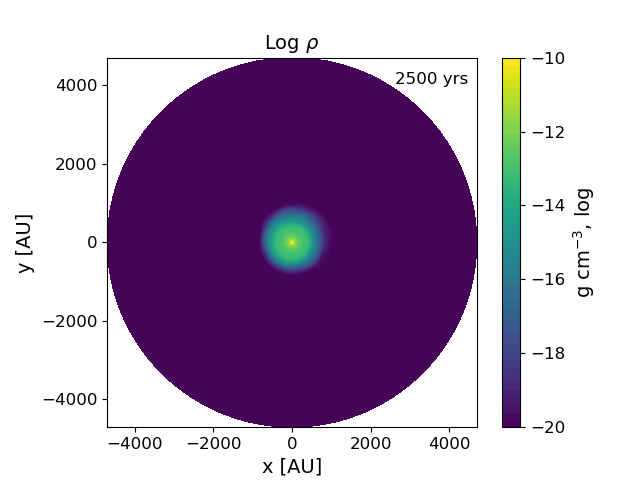}
    \includegraphics[width=\columnwidth]{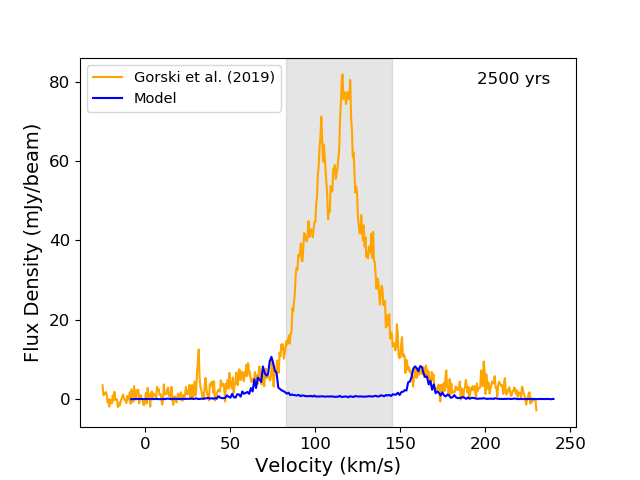}
    \includegraphics[width=\columnwidth]{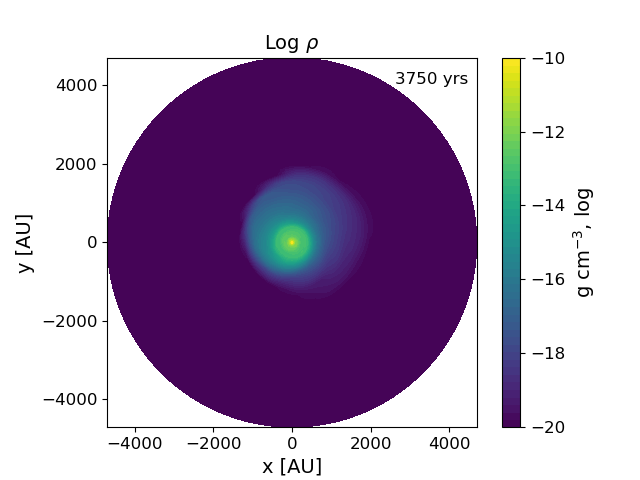}
    \includegraphics[width=\columnwidth]{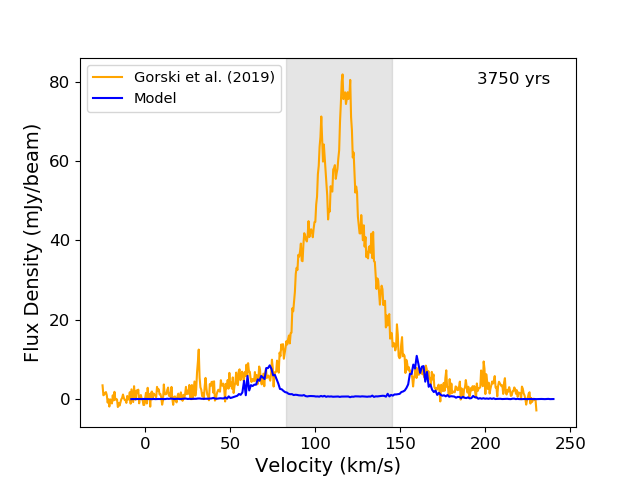}
    \includegraphics[width=\columnwidth]{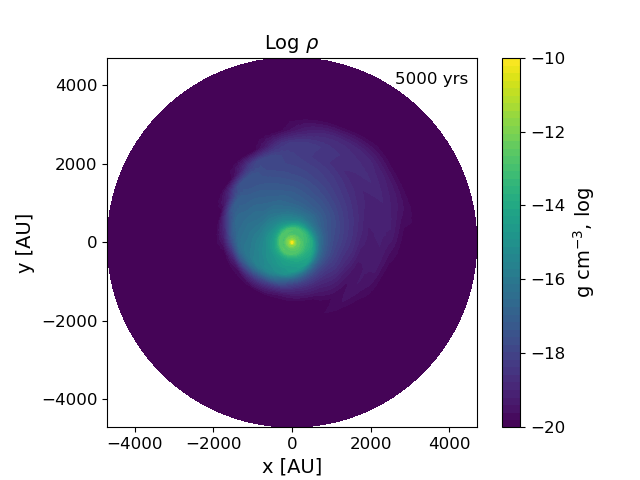}
    \includegraphics[width=\columnwidth]{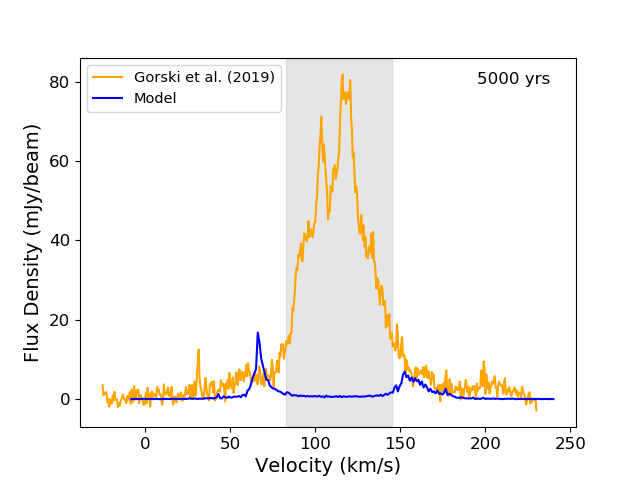}
    \caption{Time evolution of the reaction of an accretion disc around an SMS with $M_\mathrm{SMS}$ = 1000 M$_{\sun}$, $R_\mathrm{out}$ = 500 au and $M_\mathrm{disc}$/$M_\mathrm{SMS}$ = 1 per cent to a flyby rate of one stellar perturber per year. The left column shows density plots with corresponding maser spectra on the right for selected time steps. The model spectrum is shown in blue, whilst the W1 kilomaser from \citet{Gorski_2018} for comparison is plotted in orange. See caption for Fig.~\ref{fig:W1_spectrum} for the definition of different features in the maser spectrum.  }\label{fig:1000sun_500au_10sun_550peri_1yr}
\end{figure*}

\begin{figure*}
        \includegraphics[width=\columnwidth]{maser_time_SMS-mrd-hf.png}
        \includegraphics[width=\columnwidth]{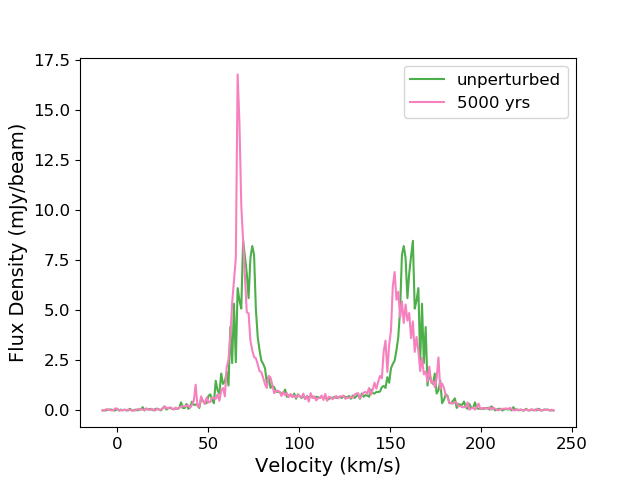}
        \caption{Time evolution of the maser spectra for SMS with parameters same as Figure~\ref{fig:1000sun_500au_10sun_550peri_1yr} (left) and Figure~\ref{fig:10000sun_500au_100sun_550peri} (right). The unperturbed maser spectrum, i.e. at time 0 yr, is plotted green and at the end of the simulation run is plotted pink. }
        \label{fig:maser_time}
\end{figure*}

Figure~\ref{fig:1000sun_500au_10sun_550peri_1yr} (left column) shows density plots for three time steps for run SMS-mrd-hf with small, low-mass accretion disc and high perturber frequency. A slow, steady and smooth dispersion of the disc is well visible. Some gas is being spread out over 2000~au distance within 5,000 years. Spiral features develop, but the main part of the original disc remains intact. In the right column we show the comparison of the modelled H$_2$O maser spectra for each time step with the W1 kilomaser \citep{Gorski_2018} in orange. The system velocity was added to the modeled velocities for direct comparison to the observed spectrum. The 'high-velocity' features are clearly visible on each side, with comparable flux value but slightly closer to the systemic velocity than the W1 kilomaser. Those values are approximately 74 and 160~km~s$^{-1}$ at timestep 2,500 years (top row). Using the Eq.~\ref{eq:outerrad} with $\Delta \nu \sim$ 43~km~s$^{-1}$, gives a radius for the maser spots of $\sim$ 480~au. As expected, this is the largest radius in the model disc where the gas temperature is still high enough for water masers. As the disc evolves with time the 'high-velocity' feature on the left becomes broader with lower flux (middle row, Figure~\ref{fig:1000sun_500au_10sun_550peri_1yr}) between 60 to 75~km~s$^{-1}$, giving the radius of the maser spots at around 380. It can also be noticed that the the 'high-velocity' feature at around 66~km~s$^{-1}$ at the time step of 10,000 years has a slightly higher flux to the feature at 153~km~s$^{-1}$. Looking at the corresponding density plot it can be observed that a spiral arm has formed, moving clockwise and creating a long, velocity-coherent region near $\sim$ 350~au, whilst the maser spots corresponding to the lower-flux feature at 153~km~s$^{-1}$ are located at the radius of $\sim$ 650~au. Additionally it can be noticed from the time evolution of the maser spectra in the same figure that the spikes of the 'high-velocity' features are moving slightly inwards or outwards, depending on the location of the spiral arm, visible in the corresponding density plots. That effect is clearly visible on the left plot in Figure~\ref{fig:maser_time} where we compare the unperturbed maser spectrum, i.e. at the very beginning of the simulation run, to the spectrum produced at 5,000 years.

\subsection{Disc stability for a 10,000~M$_{\sun}$ star}
\begin{figure*}
    \includegraphics[width=\columnwidth]{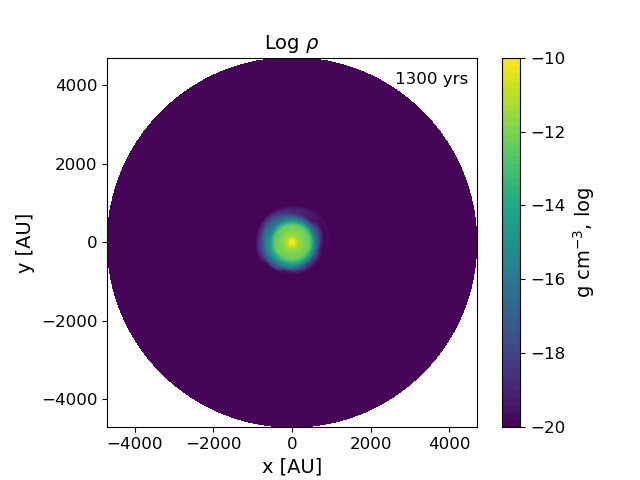}
    \includegraphics[width=\columnwidth]{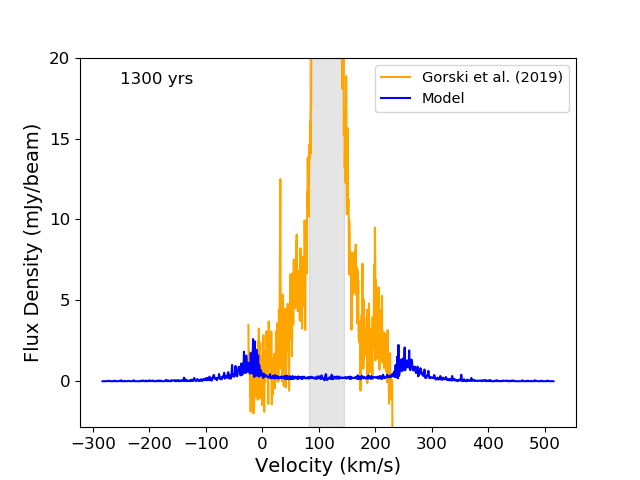}
    \includegraphics[width=\columnwidth]{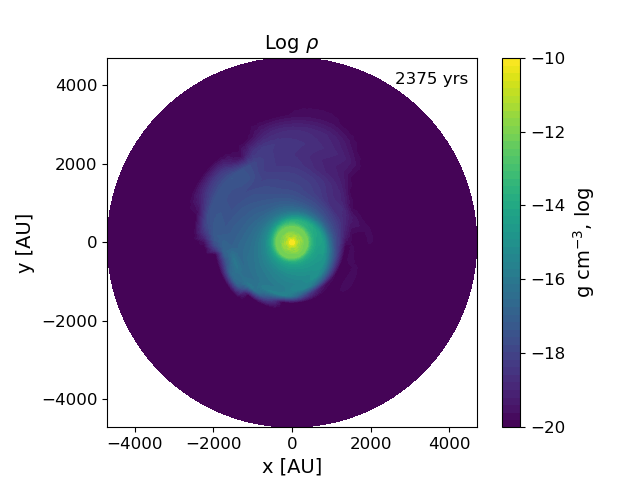}
    \includegraphics[width=\columnwidth]{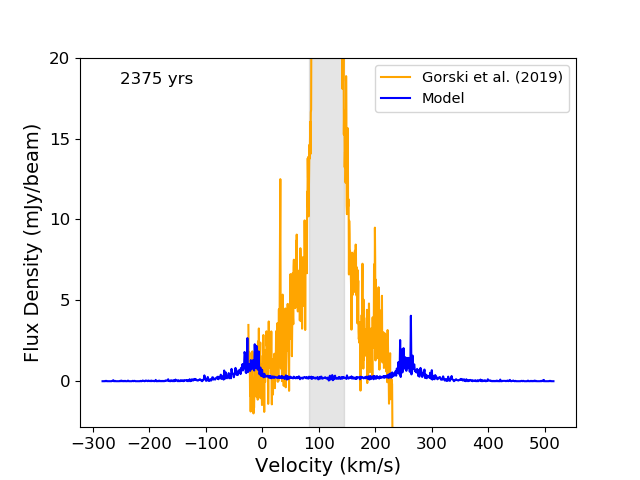}
    \includegraphics[width=\columnwidth]{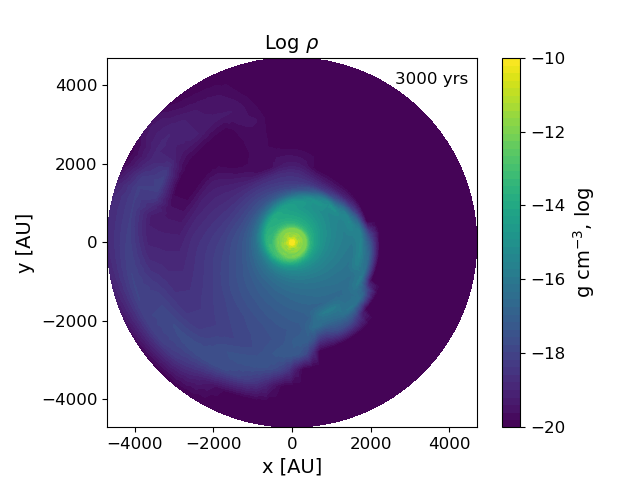}
    \includegraphics[width=\columnwidth]{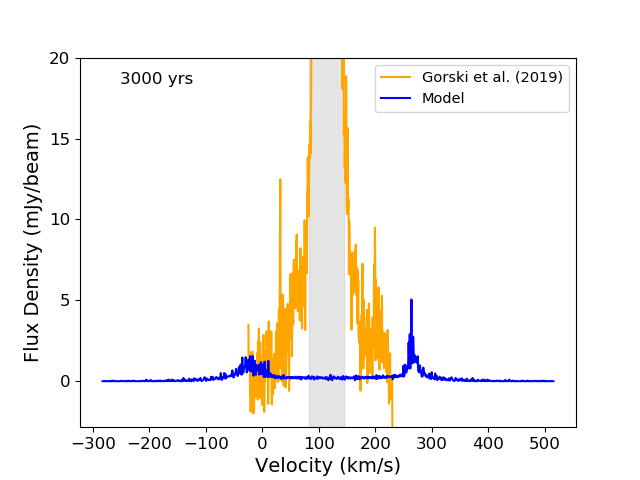}
    \caption{Same as Fig.~\ref{fig:1000sun_500au_10sun_550peri_1yr} for run SMS-Mrd-hf, which has the same parameters as the simulation in Fig. 2 except the SMS mass is now 10,000~Msun. See caption for Fig.~\ref{fig:W1_spectrum} for the definition of different features in the maser spectrum.}
    \label{fig:10000sun_500au_100sun_550peri}
\end{figure*}

Comparing the results for the case of a 10,000~M$_{\sun}$ SMS (simulation run SMS-Mrd-hf)  in Figure~\ref{fig:10000sun_500au_100sun_550peri} to the lower mass one, keeping the other parameters the same, we notice that the disc spreads out faster, with more distinct spiral arms and more mass reaching a larger radius. It is hard to establish what maximum distance the mass could reach since we are limited by the computational grid (up to 5000 au). Since the gas rotates much faster than in the previous case, the velocity range in the maser spectrum (right column in Fig.~\ref{fig:10000sun_500au_100sun_550peri}) had to be extended to 800~km~s$^{-1}$ in order to see the 'high-velocity' features, which appear in the first snapshot (top row in Fig.~\ref{fig:10000sun_500au_100sun_550peri}), symmetrically located at around -13 and 240~km~s$^{-1}$. Those values gives $\Delta \nu \sim$ 126~km~s$^{-1}$, with Eq.~\ref{eq:outerrad} we get the radius of maser spot of $\sim$ 550~au, slightly inwards of the $T$ = 300~K limit above which we allow maser emission. This, again, reflects the quick build-up of structure in the gas distribution. As the disc evolves with time, the mass and spiral arms spread out more (middle row in Figure~\ref{fig:10000sun_500au_100sun_550peri}) and the 'high-velocity' features in the corresponding maser spectrum move slightly outwards from -13  to -26~km~s$^{-1}$ and from 240 to 263~km~s$^{-1}$, leading to the maser spots being now located at the radius of 424~au, therefore slightly closer to the central star. At timestep of 3,000 years (bottom row) the 'high-velocity' feature on the left of the systemic velocity has broadened its peak whilst the spike on the right has increased its flux due to the movement of the spiral arm and increasing column density along the line of sight, as clearly seen on the right plot in Figure~\ref{fig:maser_time}.

\subsection{Effects of changing disc mass, radius and flyby rate}
The results, especially with same parameters but varying flyby rates have a moderate impact on how the density plots and model maser spectra look like. Similarly we do not see strong differences in simulations where the mass or radius of the disc is varied.

\begin{figure*}
    \includegraphics[width=\linewidth]{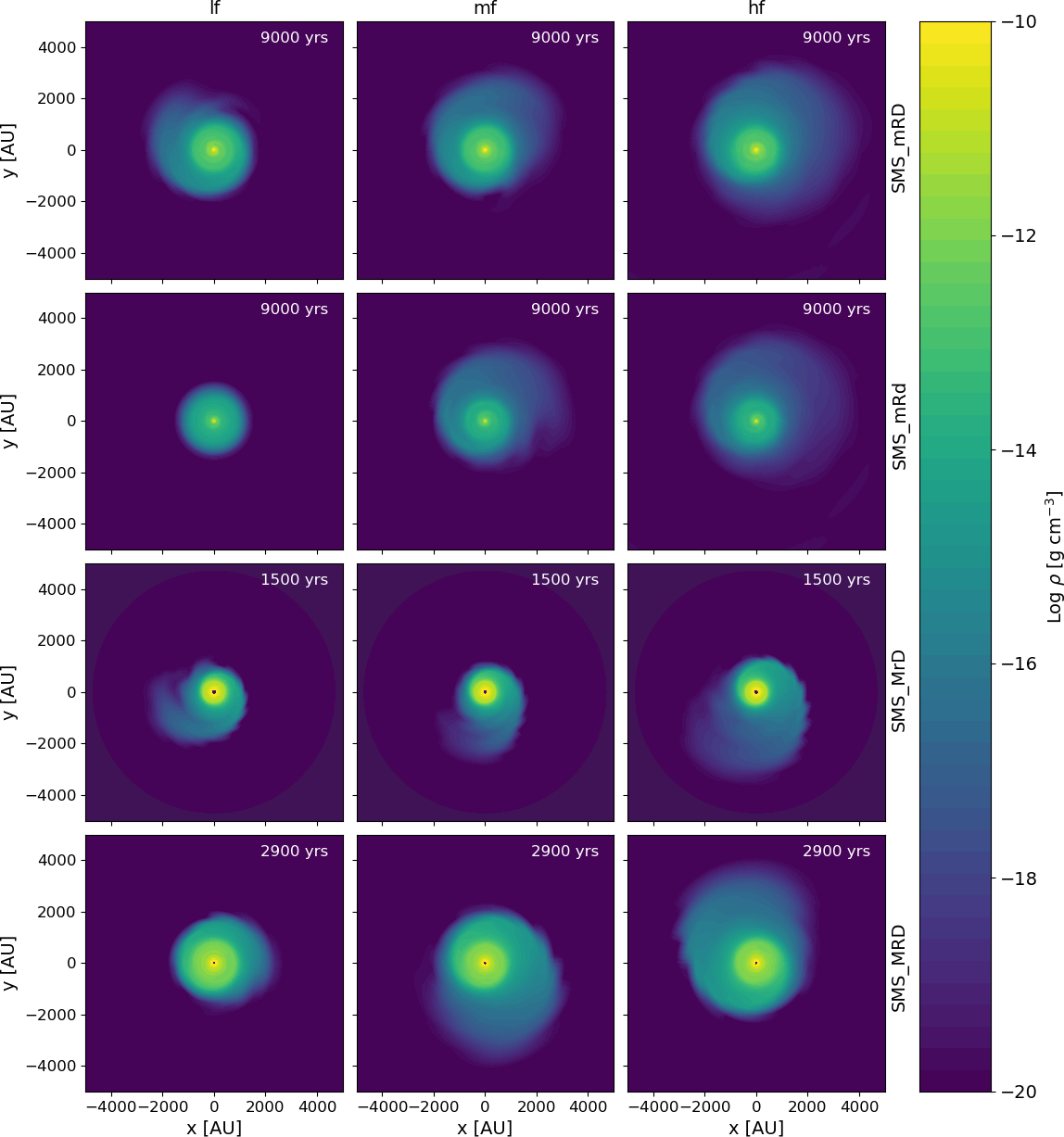}
    \caption{Comparison of density plots for simulations with different parameters. Each columns represents different flyby rates, left: 1 per 100 years (lf), middle: 1 per 10 years (mf), right: 1 per 1 year (hf). Simulation run in top row: SMS-mRD, 2nd row: SMS-mRd, 3rd row: SMS-MrD and bottom row: SMS-MRD.}
    \label{fig:simulations_comparison}
\end{figure*}

Figure~\ref{fig:simulations_comparison} shows comparison of the density maps of the snapshot for selected simulation runs for similar stages of disc evolution. In the simulations with higher disc mass (top row, simulation
run SMS-mRD) compared to the lower one (2nd row, simulation run SMS-mRd) it is noticeable that more disc mass is being spread out over larger distances, especially for the low flyby rate (left column). Since the gravitational acceleration is only proportional to the potential of the central star, the amount of mass in the disc does not influence how fast the gas rotates. There is, however, a hydrodynamic interaction of the disc with the background gas, which allows a disc with more inertia to spread somewhat faster. Especially for the 10,000~M$_{\sun}$, where the disc gas rotates much faster, the spiral arm spreads out more vigorously and becomes more distinct i.e. dense (3rd and bottom rows, simulation runs SMS-MrD and SMS-MRD, respectively). This also shows that the background gas, with its chosen low value, is still of some significance for the spreading of the gas. We do not notice any substantial change in the disc with radius of 500 au (3rd row) compared to the 1000~au one (bottom row), apart from the disc spreading faster for the lower radius case. This is a consequence of the lower value of the maximum periastron distance for the perturbers' orbit. We notice that the higher flyby rates make the disc spread out faster (compare each column in Fig.~\ref{fig:simulations_comparison}, representing different flyby rates). 

\section{Discussion}\label{s:discussion}
We have simulated the stability of accretion discs around so far hypothetical supermassive stars in the dynamic environment of a young massive cluster. All our discs preserve a dense central region, and spread out at different rates, depending on the exact parameters we chose.

In the multiple stellar flybys around the supermassive star, a single 0.2~M$_{\sun}$ stellar perturber does not have any influence on the central star and its disc. This agrees with the results from \cite{Vorobyov_2020}, where with decreasing perturber mass the effect on the disc is gradually lessening. However multiple perturbers of that low mass do have consequences on the disc, for all the chosen simulations parameters (compare different columns in Fig.~\ref{fig:simulations_comparison}). Taking the case of the most destructive flyby rate of one stellar perturber per year for the 1000~M$_{\sun}$ star, where we have entered 5000 perturbers of 0.2~M$_{\sun}$ mass onto the computational domain, the total mass adds up to 1000~M$_{\sun}$. The total mass of the perturbers therefore equals the mass of the central star, the SMS. Comparing the case for the 10~M$_{\sun}$ accretion disc (Figure~\ref{fig:1000sun_500au_10sun_550peri_1yr}) with the single, equal-mass, stellar flyby seen in \citet[their Figure 2, top row]{Cuello_2018}, the overall magnitude of the effects on the accretion discs are very comparable, in particular, the disc spreads out. We also noticed some differences: the density plots of a single perturber parabolic flyby show the disc mass being pulled out as one track, often the shape is reminiscent of spiral arms, whilst multiple flybys show much smoother structure with no preferred direction of the ejected mass from the disc. This is of course an expected consequence of the isotropic bombardment of the disc by the large number of stellar perturbers hitting over a long time allowing the mass distribution to become more homogeneous over larger distances.

Including radiative transfer in our simulation would have some effect on the disc dynamics, changing opacity and therefore heating different parts of the disc at different times. This might contribute to some expansion of the disc. The sound speed, $c_\mathrm{s}$, for our simulation model is about 1.0~km~s$^{-1}$, whilst the corresponding Keplerian velocity, $v_\mathrm{k}$, ranges between $\sim$ 200~km~s$^{-1}$ and $\sim$~42.0 km~s$^{-1}$. Since $c_\mathrm{s}$ << $v_\mathrm{k}$ we would not expect the disturbances introduced by live radiative transfer to be significant.

Incorporating magnetic fields to the model, would result in the magneto-rotational instability (MRI), that would in turn lead to viscosity, disc diffusion and accretion. As a consequence the disc would spread, decrease its density and hence have a lower flux in the 'high-velocity' features of our maser model. We model a star that grows by collisions and the accretion rate by gas is assumed to be low. Hence it is implicit in the model assumption that the MRI should have small effects during the simulation time. In a realistic scenario spreading by MRI would likely be offset by the inflow of additional gas into the circumstellar environment. 

Self-gravity should not play a major role for the evolution of our model discs as the ratio of the mass of the disc to the mass of the central star is small. Toomre's stability parameter, $Q_\mathrm{T}$, \citep[][their Equation 15]{Krause_2013} evaluated for our disc parameters is in range $\sim$ 3.4 - 9.2. Therefore the disc is stable against gravitational collapse.

The model maser spectra produced here successfully reproduce the 'clean' disc maser case, as they have two of the features mentioned before, namely, both the 'high-velocity' features \citep[compare][]{Pesce_2015}. Moreover, as we are only simulating collisionally pumped maser emissions, therefore the systemic feature is not present in our model. In order to model those features, the simulation would have to include background infrared radiation from the central source. 

To calculate the flux in the maser model we used Eq.~\ref{Kartje_fluxes} taken from \citet{Kartje_1999}. This expression is considered to be a general formula, assuming optimal collisional pumping conditions. The authors of the paper derive the equation, by considering a velocity-coherence 'box' in an edge-on disc, to which the maser emission is confined to \citep[][their Equation 9]{Kartje_1999}. The results that we obtained from the simulations are in agreement with the observations, while there is some freedom in choosing spot sizes and linewidth, which both influence the predicted flux. We regard this as a success of the model. 

The spectrum of the unperturbed disc in both simulations, seen in Fig~\ref{fig:maser_time}, is not smooth. This is due to the resolution from the computational grids for simulations and spectral modelling. This leads to substantial flux changes from cell to cell as well as between snapshot times, when the maser beam direction has moved on.

The 1000~M$_{\sun}$ simulations generally reproduce the spectral features of the W1 kilomaser better. Apart from similar fluxes, for both 'high-velocity' features,  there is a general trend for the signal to follow an increase towards the systemic velocity for the left hand side of the maser for many models. However looking at the right hand side, the model does not follow the trend of the signal of the W1 kilomaser for the models in all the cases of different parameters. The 'high-velocity' features in the W1 kilomaser look different on the respective sides of the systemic velocity. This is similar to our model spectra, which also show similar differences between the two sides. 

The model spectra for the higher mass SMS, 10,000~M$_{\sun}$ fit worse to the W1 kilomaser. The velocity range was substantially increased in order for the two 'high-velocity' features to be visible. The model maser for the higher mass of the central star hence starts to resemble spectra for AGN megamasers, in terms of the offset of those features from systemic velocity. Thus summarising the above findings, the model spectra from simulations for the $M_\mathrm{SMS}$ = 1000~M$_{\sun}$ are a better match with the W1 kilomaser spectrum from \citet{Gorski_2018}.

The H$_2$O molecule is dissociated close to the star where $T$~>~1500~K. Those conditions do not allow for the water to stay molecular, hence the 22 GHz H$_2$O lines cannot be mased. Similarly the required water level population inversion does not occur at a larger distances, where temperature remains below 300~K. The model maser spots move inwards and outwards within that distance range, depending on the movement of the spiral arm, and its density. The higher the disc mass, the more it spreads out and the denser the spiral arm is. 

The brightest maser spots are located roughly in the distance between 350-650~au, for all the models and this would easily accommodate bloated SMS models that could have radii of tens of au. The 'high-velocity' features do not appear closer to the inner radius and the center, as the temperature increases closer to the center and decreases with the distance approaching the outer radius. We used temperature profile of $T(r)$ = $T_\mathrm{in} \left(R/R_\mathrm{in}\right)^{-q}$, where $T_{\mathrm{in}}$ indicates the effective temperature at the inner radius, $R_\mathrm{in}$ and $q$ denotes the power-law index; in our simulations  $q$ = 0.5. Changing the value of $q$ affects the radius of the maser spots, e.g., $q$ = 0.75 \citep{Dullemond_2010,Vural_2014} would move the brightest maser emission towards the inner radius. However, we have taken care to chose our temperature and radius for the star very conservatively. We believe it is unlikely that the mass of the central object would be much lower than our estimate. 

We assumed that the disc is edge-on to the observer, but if the disc is inclined then the maser spectrum might start showing slight changes, i.e. decrease in the velocity and flux values, especially of the 'high-velocity' features. As the edge-on disc is 90${\degr}$, if the disc would then be observed at some small angle i.e. 85${\degr}$, then the flux and velocity values in the maser spectrum will decrease. The inclination is well constrained by the fact that we see maser spots, and the observed velocities are hardly affected by a quite possible small deviation from an 'edge-on' disc.

\begin{figure*}
    \includegraphics[width=\columnwidth]{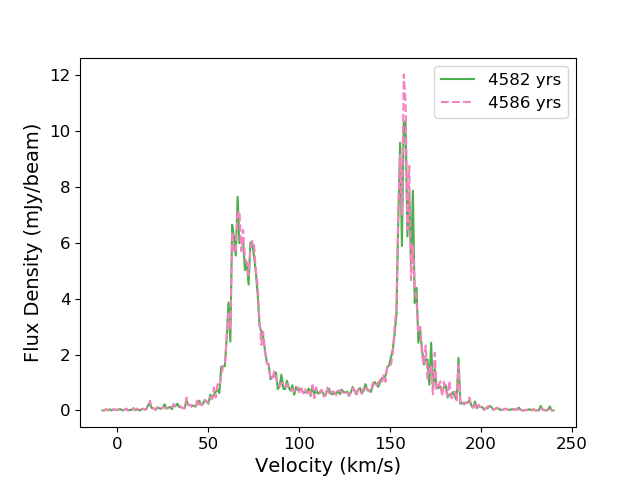}
    \includegraphics[width=\columnwidth]{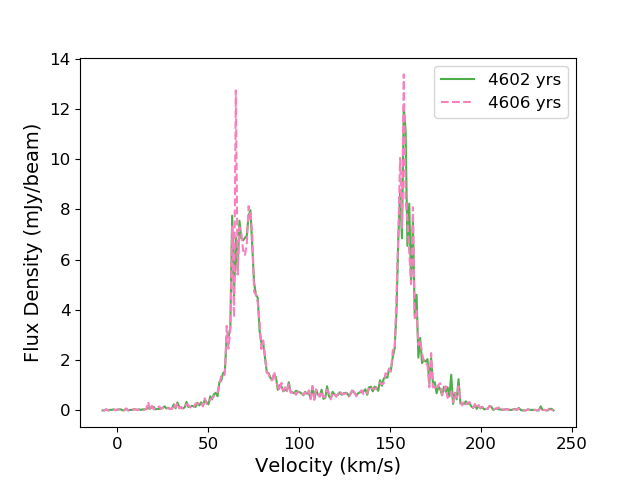}
    \caption{Examples of the maser spectrum evolution for a time interval of four years for simulation run SMS-mrd-hf. Green lines represent the model maser spectrum at the beginning of the chosen time interval and the pink dashed lines show the spectrum at the end of the four year time interval. Both time intervals have been chosen to present the biggest visible change in the spectrum during the simulation time. This is to show what changes could potentially be expected from repeated observations of the W1 kilomaser, presented in~\protect\cite{Gorski_2018}.}
    \label{fig:SMS_yr_output}
\end{figure*}

The presence of the dense spiral and their movement around the computational grid causes 'high-velocity' features to modulate the peak heights and move them inwards or outwards in the model maser spectra. The variability is in particular a consequence of the beamed nature of maser emission. Figure~\ref{fig:SMS_yr_output} displays two samples of maser spectrum evolution within chosen periods of 4 years for simulation SMS-mrd-hf, where either one of the 'high-velocity' features has visibly increased its flux or has slightly moved outwards from the systemic velocity. Hence it would be beneficial to observe the W1 kilomaser with the Very-Long-Baseline Interferometry (VLBI) to obtain more data for another epoch in order to see if there is a slight displacement for the peaks of the 'high-velocity' features.  Those findings could potentially confirm the presence of the dense spiral arms moving around the central massive object in the forming superstar cluster in NGC 253.

We note that \citet{Levy_2022} presented new 350 GHz dust emission observations of the W1 host cluster. These observations seem to indicate that the source 11 from \citet{Leroy_2018} splits up into four different objects. In their analysis the largest subsystem is 20,000~M$_{\sun}$. This substructure could also relate to a forming superbubble where the dust is beginning to be displaced from the host cluster.  

The 'high-velocity' peaks in the W1 kilomaser are observed at 31.6 and 199~km~s$^{-1}$. The model spectra for $M_\mathrm{SMS}$ = 1000~M$_{\sun}$ show a smaller velocity spread between the peaks whilst for $M_\mathrm{SMS}$ = 10,000~M$_{\sun}$ it is larger. We can already conclude that in order for the model peaks to appear at the same velocity values as the W1 kilomaser, the $M_\mathrm{SMS}$ value needs to be set between these two extreme values, if the disc temperature declines in a similar way as we assumed in our simulations. We mentioned before that the maser spots for both cases are then located between 380-550~au, with the average value, $R_\mathrm{out}$ = 465~au. The mean value of velocity spread between systemic and peak velocity of W1 kilomaser spectrum is $\Delta \nu$ = 83~km~s$^{-1}$. Using those values and the Eq.~\ref{eq:outerrad}, we estimate that $M_{\mathrm{SMS}} \approx$ 4000~M$_{\sun}$ would best reproduce features of the W1 kilomaser.

\section{Conclusions}\label{s:conclusions}
We have shown that the W1 kilomaser is a clean disc maser, according to the definition of \cite{Pesce_2015}, showing all three expected maser line systems (compare Sect.~\ref{s:W1}). Flux and velocity spread are intermediate between massive star and AGN masers. The flux expected for a model SMS accretion disc agrees well with the one of the W1 kilomaser.

Using a range of plausible parameters for flybys and collisions of perturbers with SMS, we show that in all cases an accretion disc survives for the entire simulation time and is able to produce prominent water masers with similar strength as observed in W1.

The results showed that maser models produced for the simulations where $M_\mathrm{SMS}$ = 1000~M$_{\sun}$ are a better match for the W1 kilomaser from \citet{Gorski_2018}. The model exhibits two 'high-velocity' features with similar fluxes and the left-hand side of the model spectrum shows a similar trend of the flux with the velocity. For larger stars with mass of 10,000~M$_{\sun}$ the 'high-velocity' peaks extend their velocity values beyond the observed spectrum, starting to resemble megamasers from AGNs. Hence we can conclude that, within our assumptions, the SMS with 10,000~M$_{\sun}$ is too high a mass for the W1 kilomaser. 

We included plausible upper and lower limits for the number of perturbers entering the grid at a given time, the mass of the accretion disc and its outer radius as well as the mass of the central supermassive star. The simulations show the expected variations, e.g., for the rate the disc spreads out, but overall, the model maser spectra did not show significant differences between them. 

We focused on the most destructive case for the disc i.e. the flyby rate of one stellar perturber per year, as well as including only prograde and coplanar encounters. We conclude that dynamic flybys would not impact any SMS accretion disc too strongly and hence would not prevent the formation of maser spots.

Summarising the results presented here, we can confirm that the results support the hypothesis for a supermassive star being present in the forming massive star cluster, and potentially being able to produce the best match to the observed maser spectrum of the W1 kilomaser obtained with an SMS around 4000~M$_{\sun}$.

\section*{Acknowledgements}

We thank the anonymous reviewer for useful comments and Mark Gieles, Corinne Charbonnel and D. J. B. Smith for helpful comments on an earlier version of the manuscript. This work was supported by Science and Technology Facilities Council (ST/V506709/1).

\section*{Data Availability}

The simulation data presented in  this article will be shared on reasonable request to the corresponding author.



\bibliographystyle{mnras}
\bibliography{example} 







\bsp	
\label{lastpage}
\end{document}